\newcommand{\CLASSINPUTtoptextmargin}{1in}
\documentclass[conference,10pt]{IEEEtran}

\IEEEoverridecommandlockouts
\usepackage{subcaption}
\usepackage{cite}
\usepackage{graphicx}
\usepackage{graphics}
\usepackage[cmex10]{amsmath}
\usepackage{epsfig,epic}
\usepackage{array}
\usepackage{epstopdf}
\usepackage{fancyhdr}
\usepackage{verbatim}
\usepackage{multirow}
\usepackage{multicol}
\usepackage{ctable}
\usepackage{amsfonts}
\usepackage{soul}
\usepackage{dblfloatfix}

\usepackage{makecell}
\usepackage{cellspace} %
\begin{document}
\title{\huge{Machine Learning based Interference Whitening in 5G NR MIMO Receiver}}
	\author{\IEEEauthorblockN{Shailesh Chaudhari, HyukJoon Kwon}
\vspace{-10mm}
}

\maketitle

\begin{abstract}
We address the problem of computing the interference-plus-noise covariance matrix from a sparsely located demodulation reference signal (DMRS) for spatial domain interference whitening (IW). The IW procedure is critical at the user equipment (UE) to mitigate the co-channel interference in 5G new radio (NR) systems. A supervised learning based algorithm is proposed to compute the covariance matrix with goals of minimizing both the block-error rate (BLER) and the whitening complexity. A single neural network is trained to select an IW option for covariance computation in various interference scenarios consisting of different interference occupancy, signal-to-interference ratio, signal-to-noise ratio, modulation order, coding rate, etc. In interference-dominant scenarios, the proposed algorithm computes the covariance matrix using DMRS in one resource block (RB) due to the frequency selectivity of the interference channel. On the other hand, in noise-dominant scenarios, the covariance matrix is computed from DMRS in entire signal bandwidth. Further, the proposed algorithm approximates the covariance matrix into a diagonal matrix when the spatial correlation of interference-plus-noise is low. This approximation reduces the complexity of whitening from $\mathcal{O}(N^3)$ to $\mathcal{O}(N)$ where $N$ is the number of receiver antennas. Results show that the selection algorithm can minimize the BLER under both trained as well as untrained interference scenarios.
\end{abstract}
\IEEEpeerreviewmaketitle


\section{Introduction}
In recent years, 5G wireless networks are becoming operational worldwide in order to meet high data-rate demands and support a wide range of services. 
The user equipment (UE) in the 5G network is required to employ multiple-input, multiple output (MIMO) technology to support high data rates \cite{andrews2014}. Further, due to a dense deployment of the base stations and high frequency reuse factor in the 5G networks, it is inevitable that the user equipment (UE) need to efficiently tackle co-channel interference (CCI) in the downlink reception \cite{nam2014}. A low-complexity approach to handle the interference in a MIMO receiver is to treat it as a colored Gaussian noise and apply spatial domain interference whitening (IW) \cite{Venkatesan2004,Tse2005}. This approach does not require the knowledge of interference channel.

Spatial domain IW uses the estimated interference-plus-noise covariance matrix to whiten the received signal and estimated MIMO channel matrix. The covariance matrix can be estimated from reference signals (RS) using the knowledge of transmitted pilot symbols \cite{Yu2013}. In the 5G new radio (NR) system, the covariance matrix is estimated from  PDSCH demodulation reference signal (DMRS). This is because the PDSCH DMRS experiences same interference as data symbols in PDSCH \cite{Guo2018}. 

\begin{figure}
	\centering
	\includegraphics[width=0.7\columnwidth]{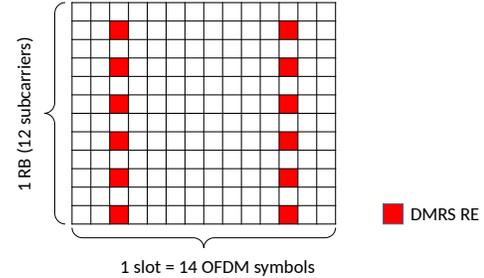}	
	\caption{{\footnotesize Illustration of sparse DMRS of type-1.}}
	\label{fig:IW_DMRS}
	\vspace{-5mm}
\end{figure}

However, there are challenges in estimating the covariance matrix from PDSCH DMRS. The first challenge is that the PDSCH DMRS can be sparsely located within the resource block (RB) \cite{Guo2018}. For example, there are only twelve DMRS resource elements (REs) available in one slot to estimate the covariance matrix as shown in Fig. \ref{fig:IW_DMRS} when the base station transmits two symbols of DMRS configuration type-1 \cite{3gpp2018_38211}. Due to fewer samples, the variance of the estimated sample covariance is high indicating that the estimated covariance is not robust \cite{Garcia2008}. To improve the estimation quality, the covariance matrix can be estimated by averaging over DMRS in all RBs in the entire signal bandwidth, provided the interference is absent or its power is low in those RBs. However, the receiver does not have a prior knowledge of the interference occupancy or power  to make a decision whether or not to use the DMRS in the entire signal bandwidth for covariance estimation.

The second challenge is the high complexity of whitening which involves Cholesky factorization of the covariance matrix followed by the inverse of a lower triangular matrix \cite{Yu2013}. The overall complexity of this procedure is $\mathcal{O}(N^3)$ where $N$ is the number of receiver antennas \cite{Boyd2004}. If the interference-plus-noise is spatially uncorrelated, then the complexity of the whitening can be reduced from $\mathcal{O}(N^3)$ to $\mathcal{O}(N)$ without any loss in performance by approximating the covariance matrix with a diagonal matrix and setting the non-diagonal terms to zero. However, in the absence of any prior knowledge of the interference power, it is not straightforward to decide whether the interference-plus-noise is spatially uncorrelated or not by observing only a few PDSCH DMRS samples.

Recent works in the literature have employed machine learning to mitigate the effects interference in MIMO receiver \cite{shlezinger2020,shlezinger2021,park2021}. However, the methods developed in \cite{shlezinger2020,shlezinger2021} tackle inter-stream interference among different MIMO streams or layer. Those methods are not applicable in CCI scenarios where the interference channel is unknown. Further, the paper \cite{park2021} proposed a reinforcement learning (RL) method to compute the covariance with few DMRS samples under CCI. However, the proposed online RL technique requires the UE to learn the interference statistics at run-time. Such techniques have implicit assumption that interference occupancy and power remain constant from one slot to the next. Further, the method in  \cite{park2021} requires maintaining a large look-up table of state, action and rewards at the UE. In this work, we propose a machine learning based IW technique applicable in CCI scenarios. Our approach do not require learning interference statistics at run-time. Further, the proposed method is applicable even when the interference occupancy and power changes at each time slot. 

We employ a neural network to decide whether or not to compute the covariance matrix by averaging over entire bandwidth. The neural network also determines if the covariance matrix can be approximated with a diagonal matrix or not. The proposed method does not require any prior knowledge of the interference occupancy and power. We train a neural network based on decoding results of the serving signal under different interference scenarios and apply the same network in different scenarios.

The main contributions of this paper are summarized below.
\begin{enumerate}
\item A supervised learning based algorithm is proposed to select an IW option to compute interference-plus-noise covariance matrix,
\item The proposed algorithm selects a low-complexity  IW option and simultaneously minimizes the block-error rate (BLER),
\item A single neural network is robust even under different interference scenarios.
\end{enumerate}

\textit{Notations}: Vectors are denoted by bold, lower-case letters, e.g., $\mathbf{h}$. Matrices are denoted by bold, upper-case letters, e.g., $\mathbf{H}$. Hermitian transpose is denoted by $(.)^*$ and $diag(\mathbf{A})$ is a diagonal matrix with diagonal elements same as the diagonal elements of $\mathbf{A}$.

\section{System Model and Objective}
\label{sec:system}
We consider a $M\times N$ MIMO system model with $M$ transmit antennas and $N$ receiver antennas. Let $\mathbf{y}_{i,j,t} \in \mathbb{C}^{N\times 1}$ be the received signal vector at time slot $t$ in RE $(i,j)$, i.e., RE located at the $i$th sub-carrier and $j$th OFDM symbol in the slot. Let $\mathbb{S}_b, b=1,\cdots,B$, be the set of REs in the $b$th resource block (RB) where $B$ is the total number of RBs in the signal bandwidth. The set of RBs with interference is denoted by $\mathbb{I}$. Then, the received signal vector can be expressed as 

{\small
\vspace{-3mm}
\begin{align}
\small
\mathbf{y}_{i,j,t} = 
\begin{cases}
\mathbf{H}_{i,j,t} \mathbf{x}_{i,j,t} + \mathbf{n}_{i,j,t}, \text{if } (i,j) \in \mathbb{S}_b, b \notin \mathbb{I},\\
\mathbf{H}_{i,j,t} \mathbf{x}_{i,j,t} + \mathbf{H_I}_{i,j,t} \mathbf{x_I}_{i,j,t} + \mathbf{n}_{i,j,t}, \text{if } (i,j) \in \mathbb{S}_b, b\in \mathbb{I},
\end{cases}
\label{eq:y}
\end{align}
\vspace{-3mm}
}

\noindent where $\mathbf{H}_{i,j,t} \in \mathbb{C}^{N\times M}$ and $\mathbf{x}_{i,j,t} \in \mathbb{C}^{M\times 1}$ are the downlink MIMO channel and the transmitted signal, respectively. Further, $\mathbf{H_I}_{i,j,t}\in \mathbb{C}^{N\times M'}$ and $\mathbf{x_I}_{i,j,t} \in \mathbb{C}^{M'\times 1}$ are the downlink interference channel and interference signal, where $M'$ is the number of antennas at the interfering transmitter. Finally, $\mathbf{n}_{i,j,t} \in CN(0,\mathbf{I})$ is the noise vector. We assume that the $\mathbf{H}_{i,j,t}$ and $\mathbf{H_I}_{i,j,t}$ are scaled according to SNR and SIR assuming unit noise power at each antennas and the channels remain constant over the duration of slot $t$. Therefore, we drop subscript $t$ to simplify notations.

We define the interference-plus-noise vector $\mathbf{v}_{i,j}$ as follows:
\begin{align}
\small
\mathbf{v}_{i,j} = \mathbf{y}_{i,j} - \mathbf{H}_{i,j} \mathbf{x}_{i,j} = 
\begin{cases}
\mathbf{n}_{i,j}, \text{if } (i,j) \in \mathbb{S}_b, b \notin \mathbb{I},\\
\mathbf{H_I}_{i,j} \mathbf{x_I}_{i,j} + \mathbf{n}_{i,j}, \text{if } (i,j) \in \mathbb{S}_b, b\in \mathbb{I},
\end{cases}
\label{eq:v}
\end{align}

\begin{figure}
	\centering
	\includegraphics[width=\columnwidth]{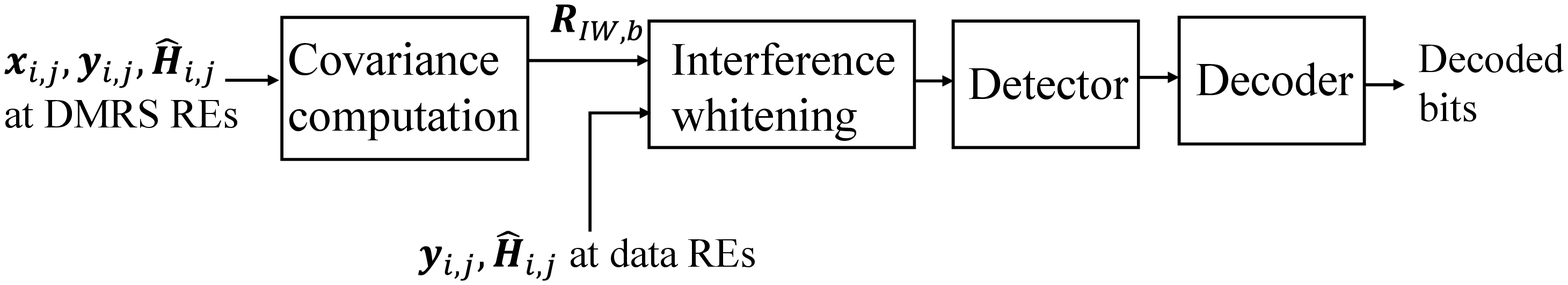}	
	\caption{{\footnotesize Block diagram of receiver with interference whitening.}}
	\label{fig:IW_system_model}
	\vspace{-5mm}
\end{figure}

Assuming that the noise and interference signal has zero mean, the ideal interference-plus-noise covariance matrix for the $b$th RB is $\mathbf{R}_{IW,b}=\mathbb{E}[\mathbf{v}_{i,j}\mathbf{v}^*_{i,j}]$. In order to whiten the signal in the $b$th RB, first the Cholesky factorization $\mathbf{L}_b$ of $\mathbf{R}_{IW,b}$ is computed such that $\mathbf{R}_{IW,b} = \mathbf{L}_{b} \mathbf{L}^*_{b}$. Then, the received signal vector $\mathbf{y}_{i,j}$ and the estimated channel matrix $\mathbf{\hat{H}}_{i,j}$ are pre-multiplied by the inverse of $\mathbf{L}_b$ as follows:
\begin{align}
\mathbf{y}^{(w)}_{i,j} = \mathbf{L}^{-1}_{b}\mathbf{y}_{i,j}, 
\mathbf{\hat{H}}^{(w)}_{i,j} = \mathbf{L}^{-1}_{b}\mathbf{\hat{H}}_{i,j}, (i,j) \in \mathbb{S}_b
\label{eq:whiten}
\end{align}
where $\mathbf{y}^{(w)}_{i,j}$ and $\mathbf{\hat{H}}^{(w)}_{i,j}$ are whitened signal vector and channel matrix, respectively. Following the whitening operation, the detector computes log-likelihood ratios from $\mathbf{y}^{(w)}_{i,j}$ and $\mathbf{\hat{H}}^{(w)}_{i,j}$ and the decoded bits are obtained from the decoder as shown in Fig. \ref{fig:IW_system_model}. Let $c$ indicate the cyclic redundancy check (CRC) flag indicating whether the block is decoded successfully ($c=1$) or not ($c=0$). Let $P_e=\Pr\{c=0\}$ be the probability of decoding error or block error rate (BLER) under different channel realizations. The objective of this work is to select $\mathbf{R}_{IW,b}$ in order to minimize $P_e$ under different interference scenarios while keeping the whitening complexity low.

\section{Proposed approach}
\label{sec:proposed_approach}
In order to obtain the whitening matrix $\mathbf{R}_{IW,b}$, we first compute an initial estimate $\mathbf{\hat{R}}_{D,b}$ from DMRS in the $b$th RB. We define $\mathbb{S}_{D,b} \subset \mathbb{S}_b$ such that $\mathbb{S}_{D,b}$ is the set of DMRS REs where the transmitted DMRS are known at the receiver. The initial estimate of the covariance matrix can be computed as follows:

{\small
\vspace{-4mm}
\begin{align}
\hat{\mathbf{R}}_{D,b} = \frac{1}{|\mathbb{S}_{D,b}|} \sum_{(m,n)\in \mathbb{S}_{D,b}} \mathbf{\hat{v}}_{m,n}\mathbf{\hat{v}}_{m,n}^*,
\label{eq:R_dmrs}
\end{align}
\vspace{-3mm}
}

\noindent where $\mathbf{\hat{v}}_{m,n}=\mathbf{y}_{m,n} - \mathbf{\hat{H}}_{m,n} \mathbf{x}_{m,n}$, $|\mathbb{S}_{D,b}|$ is the cardinality of set $\mathbb{S}_{D,b}$, $\mathbf{\hat{H}}_{m,n}$ is the estimated channel at DMRS RE $(m,n)$ and $\mathbf{x}_{m,n}$ is the transmitted DMRS pilot.

We consider three options to compute $\mathbf{R}_{IW,b}$ from $\mathbf{\hat{R}}_{D,b}$. One obvious option is to directly  $\mathbf{\hat{R}}_{D,b}$ for whitening, i.e., $\mathbf{R}_{IW,b}=\mathbf{\hat{R}}_{D,b}$. The complexity of whitening with this option is $\mathcal{O}(N^3)$ due to the Cholesky factorization and matrix inverse operations \cite{Boyd2004}. This option is suitable in an interference-dominant scenario when the correlation between interference received on two antennas is high, e.g., when the non-diagonal elements of $\mathbb{E}[\mathbf{v}_{i,j}\mathbf{v}^*_{i,j}]$ are non-negligible relative to the diagonal elements.

A lower complexity option is to use only the diagonal elements of $\mathbf{\hat{R}}_{D,b}$ for whitening, i.e., $\mathbf{R}_{IW,b}=diag(\mathbf{\hat{R}}_{D,b})$. Due to a diagonal covariance matrix, the $\mathbf{L}_b$ is also a diagonal matrix where the $n$th diagonal entry is $\mathbf{L}_b(n,n)=\sqrt{\mathbf{R}_{IW,b}(n,n)}, n=1,2,\cdots,N$. Further, $\mathbf{L}^{-1}_b$ is also a diagonal matrix with $\mathbf{L}^{-1}_b(n,n) = \frac{1}{\sqrt{\mathbf{R}_{IW,b}(n,n)}}$. Therefore, the complexity of whitening becomes $\mathcal{O}(N)$. In this case, the pre-multiplication with $\mathbf{L}^{-1}_b$ in (\ref{eq:whiten}) is equivalent to \textit{normalization} of $n$-th row of $\mathbf{y}_{i,j}$ and $\mathbf{\hat{H}}_{i,j}$ with $\sqrt{\mathbf{R}_{IW,b}(n,n)}$. This option is suitable in an interference-dominant scenario when the correlation between interference received on two antennas is low, e.g., when the non-diagonal elements of $\mathbb{E}[\mathbf{v}_{i,j}\mathbf{v}^*_{i,j}]$ are negligible relative to the diagonal elements. 

In both options above, which are suitable for interference-dominant scenarios, $\mathbf{R}_{IW,b}$ is computed from DMRS REs in the $b$th RB only. This is because the interference statistics are different in different RBs due to frequency selectivity of the interference channel. 

The third option is suitable in a noise-dominant scenario where we can approximate $\mathbf{v}_{i,j}$ as $\mathbf{v}_{i,j}\approx \mathbf{n}_{i,j}$. Assuming the noise spectrum is white and the noise power is same at each RB, we can improve the estimate of the covariance by averaging over all RBs in entire signal bandwidth as shown in (\ref{eq:option2}). In this option, the complexity of whitening is $\mathcal{O}(N)$ due to a diagonal covariance matrix.

We arrange the three options in the ascending order of complexity as follows:
\begin{enumerate}
	\item IW with normalization over RB (IWNRB):
	{\small
	\begin{align}
		\mathbf{R}_{IW,b}=diag(\mathbf{\hat{R}}_{D,b})
		\label{eq:option1}
	\end{align}		
	}
	\item IW with normalization over signal bandwidth (IWNBW):
	{\small
	\begin{align}	
		\mathbf{R}_{IW,b}=diag\left(\frac{1}{B}\sum_b\mathbf{\hat{R}}_{D,b}\right)
		\label{eq:option2}
	\end{align}	
	}
	\item IW over RB (IWRB):
	{\small
	\begin{align}	
		\mathbf{R}_{IW,b}=\mathbf{\hat{R}}_{D,b}
		\label{eq:option3}
	\end{align}	
	}
\end{enumerate}

The whitening complexity with IWNRB and IWNBW is $\mathcal{O}(N)$, while the complexity with IWRB is $\mathcal{O}(N^3)$.

\subsection{Impact of interference scenario on performance of IW options}
\label{sec:impact_scenario}
\begin{figure}[]
	\vspace{-3mm}
	\centering
	\begin{subfigure}[b]{\linewidth}
		\centering
		\includegraphics[width=\columnwidth]{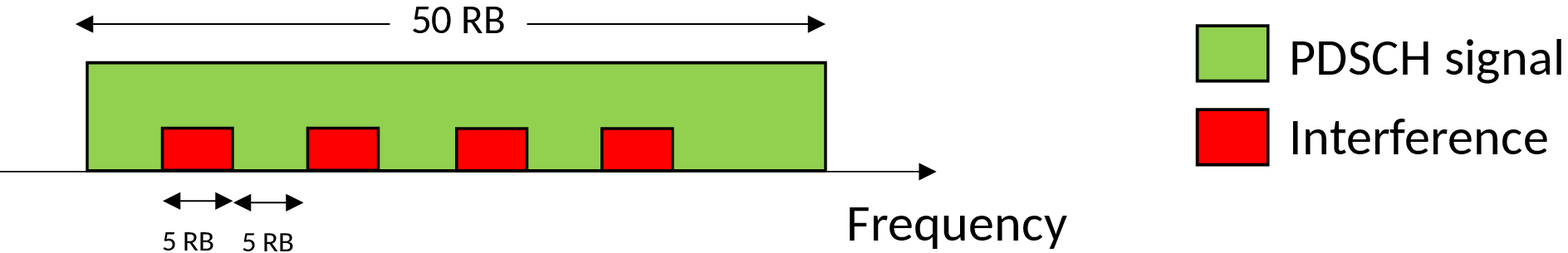}	
		\caption{{\footnotesize Interference occupancy-1: uniformly spaced interference}.}
		\label{fig:int_occ1}	
	\end{subfigure}
	\begin{subfigure}[b]{\linewidth}
		\centering
		\includegraphics[width=\columnwidth]{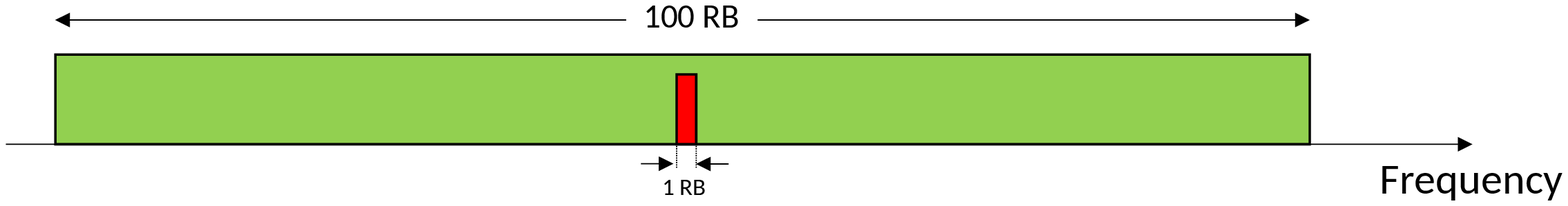}
		\caption{{\footnotesize Interference occupancy-2: concentrated interference}.}
		\label{fig:int_occ2}
	\end{subfigure}
	\caption{\footnotesize Illustration of interference occupancy.}
	\label{fig:int_occ12}
	\vspace{-3mm}
\end{figure}

\begin{figure*}[h]
	\centering
	\begin{subfigure}[b]{0.24\linewidth}
		\centering
		\includegraphics[width=\columnwidth]{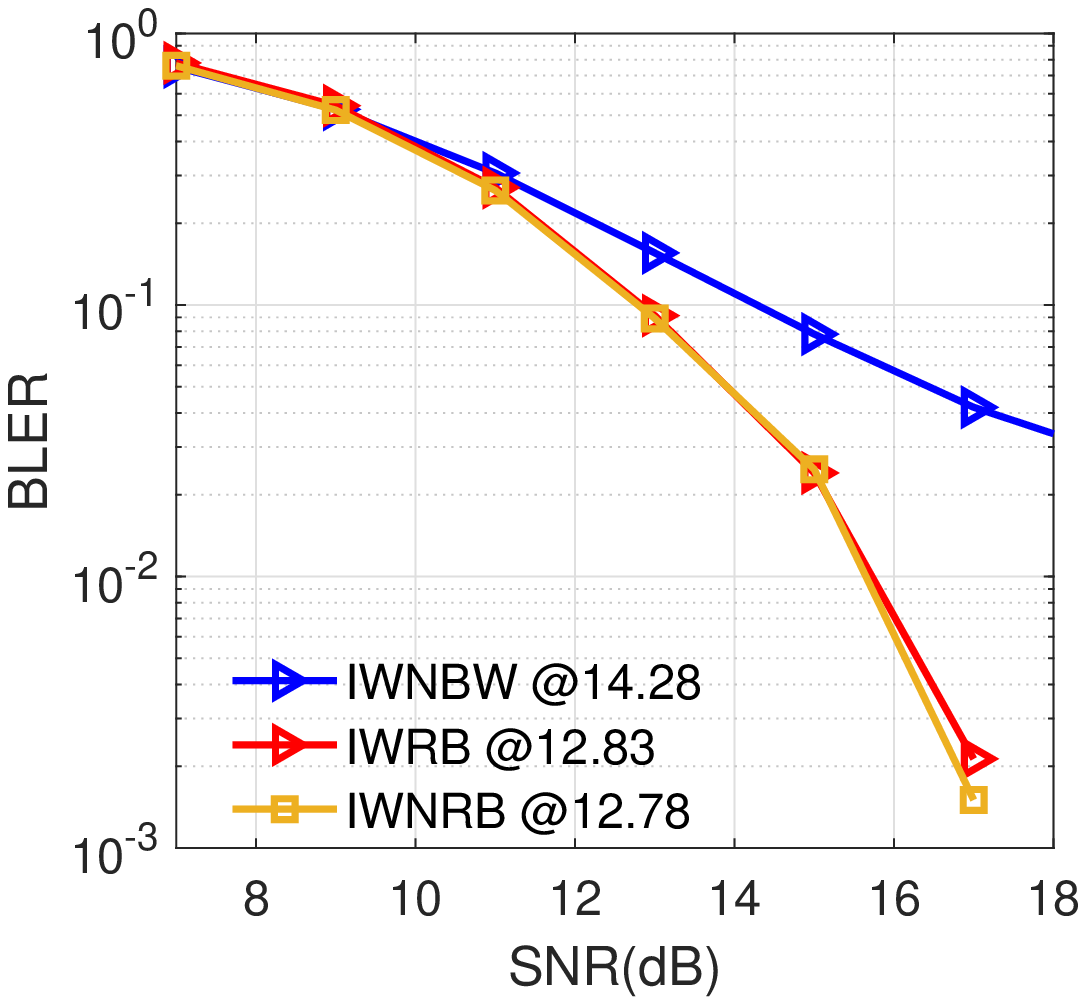}	
		\caption{{\footnotesize Occupancy-1,MCS-5,SIR=10dB}.}
		\label{fig:bler_occ1}	
	\end{subfigure}
	\begin{subfigure}[b]{0.24\linewidth}
		\centering
		\includegraphics[width=\columnwidth]{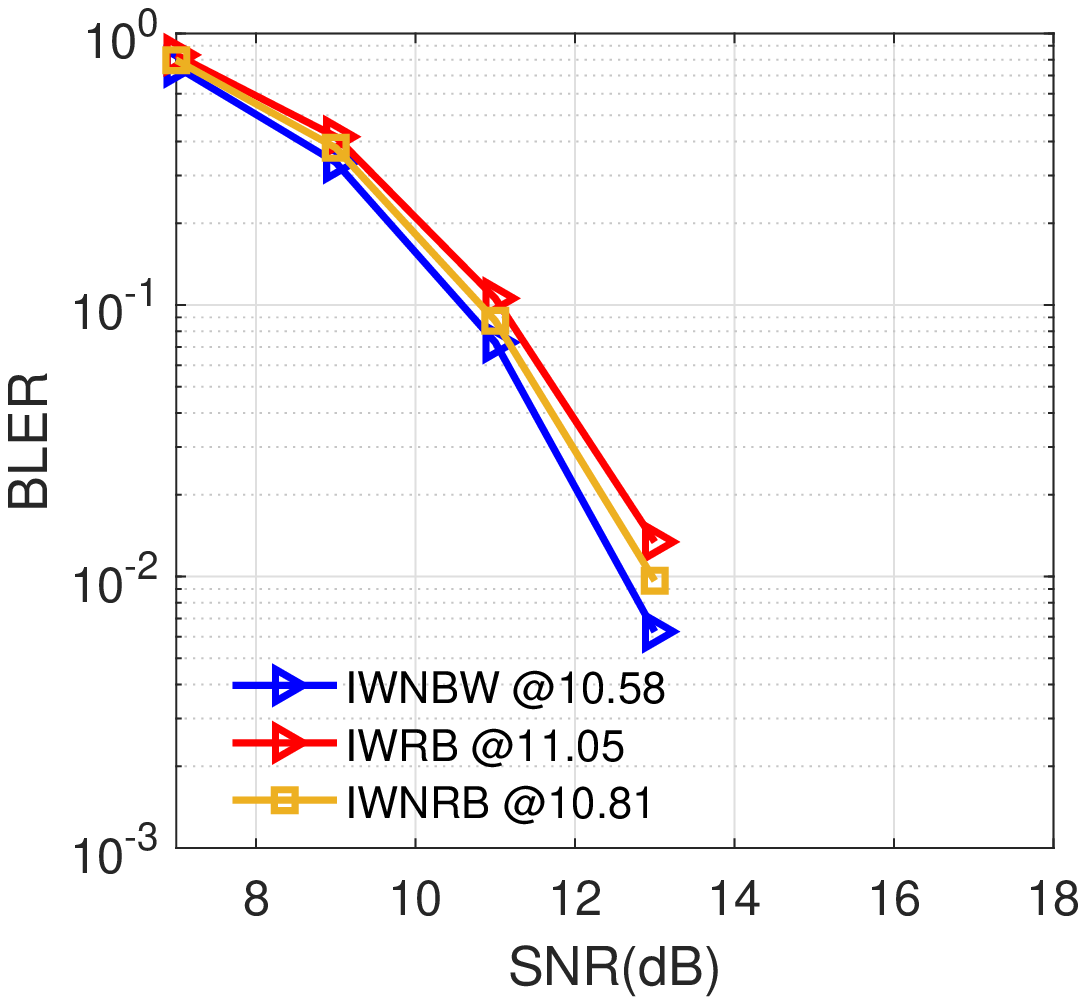}
		\caption{{\footnotesize Occupancy-2,MCS-5,SIR=10dB}.}
		\label{fig:bler_occ2}
	\end{subfigure}
	\begin{subfigure}[b]{0.24\linewidth}
		\centering
		\includegraphics[width=\columnwidth]{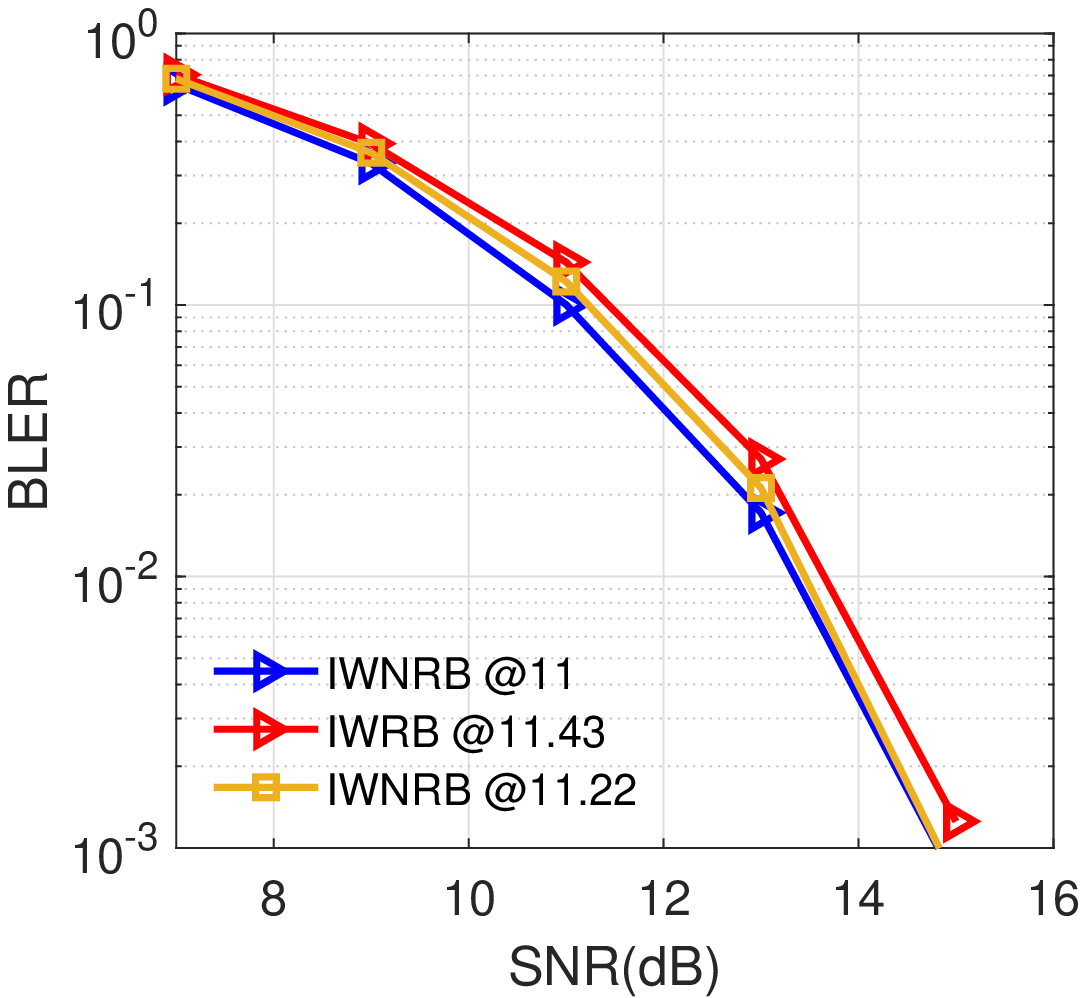}
		\caption{{\footnotesize Occupancy-1,MCS-5,SIR=30dB, }.}
		\label{fig:bler_mcs_snr1}
	\end{subfigure}
	\begin{subfigure}[b]{0.24\linewidth}
		\centering
		\includegraphics[width=\columnwidth]{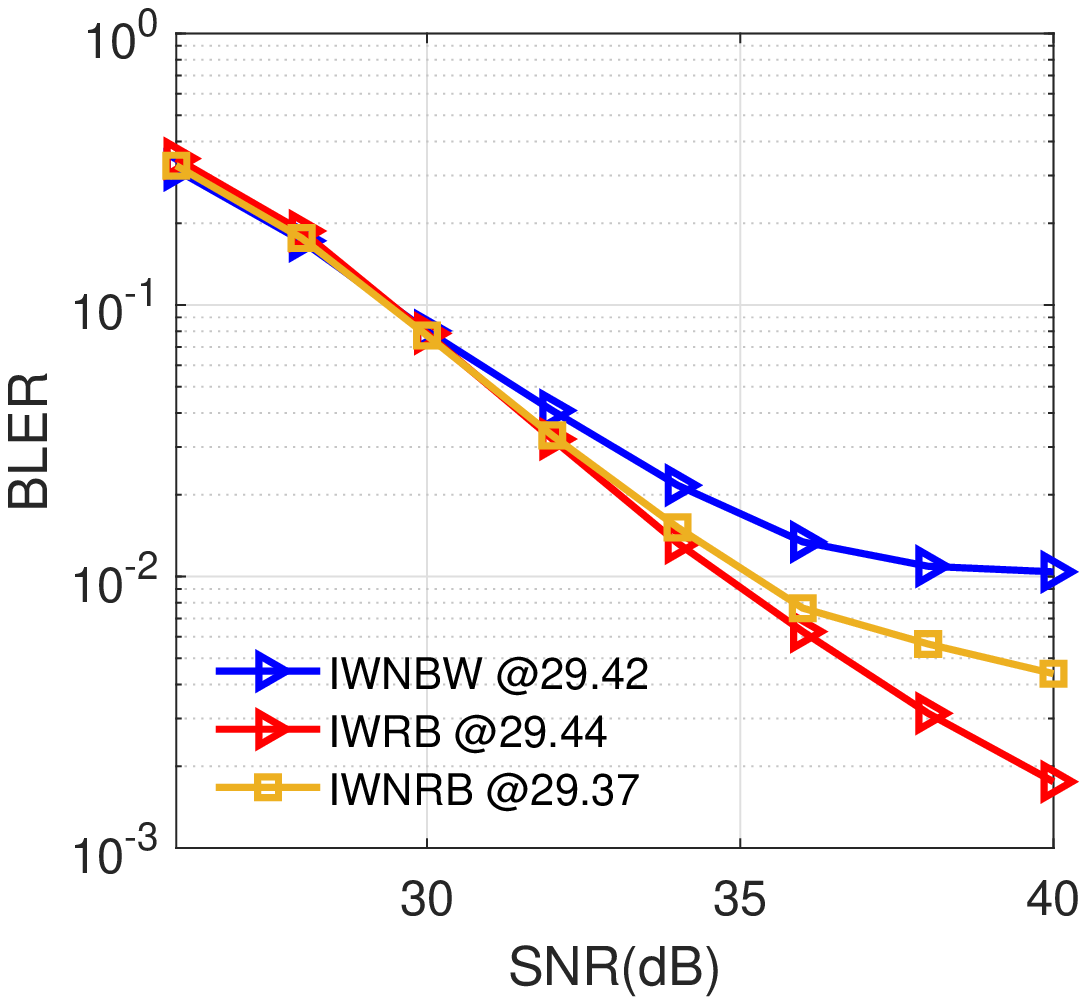}
		\caption{{\footnotesize Occupancy-1,MCS-19,SIR=30dB}.}
		\label{fig:bler_mcs_snr2}
	\end{subfigure}
	\caption{\footnotesize Effect of MCS, SIR, and interference occupancy on BLER under EPA-5 channel in 2x2 MIMO at Doppler frequency 5Hz, subcarrier spacing 15kHz, system BW=20MHz.}
	\label{fig:bler_at_diff_mcs_sir_occ}
	\vspace{-6mm}
\end{figure*}

We observe that the best IW option to minimize BLER depends on various factors such as interference occupancy in signal bandwidth, SIR, SNR, MCS as shown in Fig. \ref{fig:bler_at_diff_mcs_sir_occ}. To demonstrate the impact of these factors on BLER, we consider the interference scenarios with the following interference occupancy:
\begin{itemize}
	\item Occupancy-1: Signal of 50RB, with interference distributed throughout signal bandwidth (Fig. \ref{fig:int_occ1}), 
	\item Occupancy-2: Signal of 100RB, with interference at the center RB (Fig. \ref{fig:int_occ2}). 	
\end{itemize}

As shown in Fig. \ref{fig:int_occ12}, the interference is uniformly spaced in occupancy-1, while it is concentrated in occupancy-2. The effect of interference occupancy on the BLER is shown in Fig. \ref{fig:bler_occ1} and \ref{fig:bler_occ2} when MCS $=5$ and SIR $=10$dB. We observe that the lowest BLER is achieved with IWNRB and IWNBW options in interference occupancy 1 and 2, respectively. Finally, Fig. \ref{fig:bler_mcs_snr1}, \ref{fig:bler_mcs_snr2} demonstrate that the best option depends on MCS and SNR as well. At SIR=30dB, we can see that IWNRB is the best option for MCS-5, while IWRB and IWNRB are better options for MCS-19.

Further, we can observe the pros and cons of averaging over entire signal bandwidth in the IWNBW option. We see that the IWNBW has the lowest BLER in a  noise-dominant scenario (high SIR) as shown in Fig. \ref{fig:bler_mcs_snr1}. However, it suffers significantly in a  interference-dominant scenario (low SIR) as shown in Fig. \ref{fig:bler_occ1}. Here, we emphasize the interference occupancy is not known at the receiver. Further, the perfect knowledge of SIR and SNR is not available in the presence of interference. Therefore, we train a neural network to select appropriate IW option as described next.

\subsection{Offline Neural Network Training}
\label{sec:network_training}
The objective of the neural network is to select an appropriate IW option to minimize the BLER ($P_e$) in various interference scenarios. Interference scenario is denoted by $s$ and consists of \{Channel model, interference occupancy, SNR, SIR, MCS\}. Let $P^{(\zeta)}_{e}(s)$ be the BLER achieved when the IW option $\zeta\in \{1,2,3\}$ is applied at each time slot in a fixed interference scenario $s$. Then, the minimum BLER in scenario $s$ is denoted by $P^{(min)}_{e}(s) = \min_\zeta{P^{(\zeta)}_e}(s)$.

Let $z_{s}(\theta) \in \{1,2,3\}$ be the IW option selected by the neural network in scenario $s$, where $\theta$ is the network parameter, i.e., weights and biases of the neural network. Let $P_e(s;\theta)$ be the BLER achieved when IW option is selected by the neural network in scenario $s$. Then, the ideal the network parameter $\theta^*$ achieves $P_e(s;\theta^*) = P^{(min)}_{e}(s), \forall s$. Mathematically, the goal of network training can be stated as follows:

{\small
\vspace{-2mm}
\begin{align}
\min_{\theta}~~& \sum_{s \in \mathcal{S}} |P_e(s;\theta) - P^{(min)}_{e}(s)|,
\label{eq:objective}
\end{align}
\vspace{-3mm}
}

\noindent where $\mathcal{S}$ is the set of different interference scenarios. Labeled dataset are generated under different interference scenarios and then combined for the network training. The range of parameters used to generate the dataset is tabulated in Table \ref{table:train_dataset}. The range of SNR in the training dataset is $[SNR_{min}, SNR_{max}]$ where $SNR_{min}$ is the largest SNR with BLER $\geq 0.1$ for MCS-0 at SIR $=50$dB and $SNR_{max}$ is the smallest SNR with BLER $\leq 0.01$ for MCS-27 at SIR $=0$dB. The labeled dataset is generated by running simulation with above parameters and collecting features and labels for each slot. Each sample in the training dataset corresponds to one slot in the 5G NR transmission. The feature and label generation are explained next.

\begin{table}
	\centering
	\caption{\footnotesize{Range of scenarios in training dataset}}
	\begin{tabular}{|c|c|}
		\hline
		Parameter & Range\\
		\hline
		Channel models & EPA-5, EVA-30 \\
		\hline
		MCS & 0 to 27 from Table 5.1.3.1-2 in \cite{3gpp2018_38214} \\
		\hline		
		SIR & $\{0,10,20,30,40,50\}$dB\\
		\hline
		Interference occupancy & Occupancy-1 (uniformly spaced) and \\
		& occupancy-2 (concentrated) \\
		\hline
		SNR & $[SNR_{min}, SNR_{max}]$\\
		\hline
	\end{tabular}
	\label{table:train_dataset}
	\vspace{-4mm}
\end{table}

\subsubsection{Label Generation}
\label{sec:label_gen}
\begin{figure}
	\centering
	\includegraphics[width=\columnwidth]{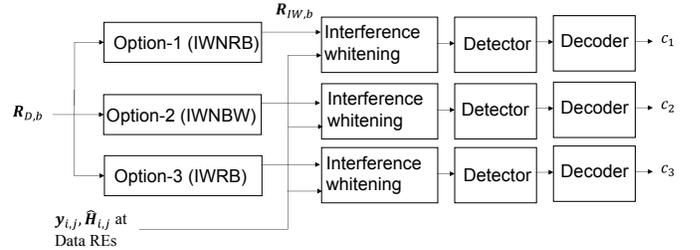}
	\caption{{\footnotesize Label generation.}}
	\label{fig:IW_label_gen}
	\vspace{-5mm}
\end{figure}

During label generation, each IW option $n\in \{1,2,3\}$ is applied in to generate $\mathbf{R}_{IW,b}$ as shown in Fig. \ref{fig:IW_label_gen}. The received signal vectors and estimated channel matrices are whitened with $\mathbf{R}_{IW,b}$ generated by each option. Finally, the CRC flag $c_n$ is collected for option $n$ at each slot. Then, the label is assigned as the lowest complexity IW option that results in successful decoding. Since the IW options are arranged in the ascending order of complexity, the lowest complexity label can be expressed as follows:
\begin{align}
z = \min_{n \in \{1,2,3\}} \{n | c_n=1\}
\end{align}

\subsubsection{Feature Generation}
\label{sec:feature_gen}
The input features for the neural network are derived from initial covariance estimate $\hat{\mathbf{R}}_{D,b}$. We define terms $R_{avg}(i), R_{max}(i),R_{min}(i)$ corresponding to the $i$-th diagonal element of $\hat{\mathbf{R}}_{D,b}$ as:
{\small
\begin{align}
\nonumber R_{avg}(i) = \frac{1}{B}\sum_{b} \hat{\mathbf{R}}_{D,b}(i,i),
\\\nonumber R_{max}(i) = \max_b \hat{\mathbf{R}}_{D,b}(i,i),
\\ R_{min}(i) = \min_b \hat{\mathbf{R}}_{D,b}(i,i)
\end{align}
\vspace{-4mm}
}

For each slot, the five features $g_1,g_2,\cdots,g_5$ shown in Table \ref{table:input_features} are generated. These feature are chosen from various candidate features based by using Mutual Information based Feature Selection (MIFS) algorithm proposed in \cite{battiti1994}. Further, as demonstrated in Section \ref{sec:impact_scenario}, the BLER for any IW option depends on the modulation order and code rate which are known at the receiver. Therefore, we utilize both modulation order and code rate in addition the features mentioned above at the input of the network.

\begin{table}
	\centering
	\caption{\footnotesize{Input features}}
	\begin{tabular}{|c|c|}
		\hline
		Feature & Description\\
		\hline
		$g_1$ & $\frac{1}{N} \sum_{i=1}^{N}R_{avg}(i)$\\
		\hline
		$g_2$ &  $\max_{i} R_{max}(i)$\\
		\hline
		$g_3$ & $\min_{i} R_{min}(i)$\\
		\hline
		$g_4$ & Modulation order\\
		\hline
		$g_5$ & Code-rate\\
		\hline
	\end{tabular}
	\label{table:input_features}
\end{table}

\begin{figure*}
	\centering
	\includegraphics[width=0.7\textwidth]{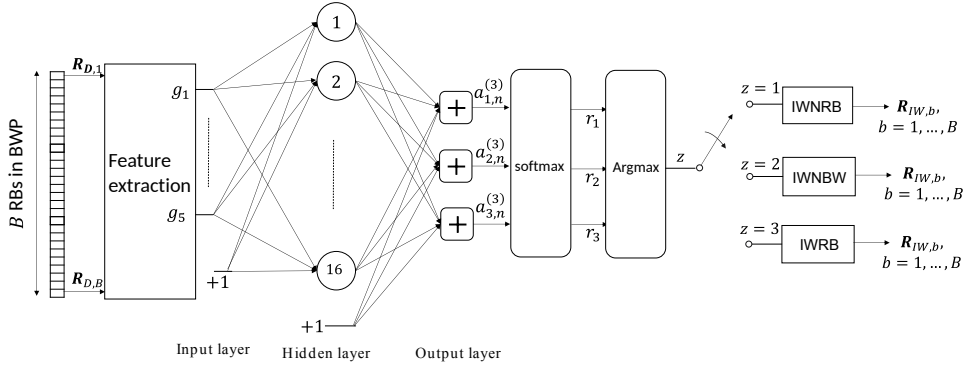}	
	\caption{{\footnotesize IW selection at each slot with neural network.}}
	\label{fig:IW_with_full_NN}
	\vspace{-4mm}
\end{figure*}

\subsubsection{Network training}
\label{sec:training}
Once features and labels are collected at different interference scenarios, a single neural network is trained using  quasi-Newton method \cite{mokhtari2015} to obtain the network parameter $\theta$. The architecture of the neural network is shown in Fig. \ref{fig:IW_with_full_NN}. It consists of 5 input features, one hidden layer with 16 hidden nodes, and 3 output nodes. We use Sigmoid activation function. The network is trained to minimize the cross-entropy cost function.

\subsection{Online IW selection}
\label{sec:online}
At the inference time, the trained parameter $\theta$ is used to select the IW option at each slot as shown in Fig. \ref{fig:IW_with_full_NN}. From the selected IW option, the covariance matrix $\mathbf{R}_{IW,b}$ is computed for each RB $b$ in the signal bandwidth.

\section{Results}
\label{sec:results}
In this section, we present results for the proposed IW selection algorithm in a $4\times 4$ MIMO system with $M=N=M'=4$. The algorithm in evaluated assuming the base station transmits two DMRS configuration type-1 as depicted in Fig. \ref{fig:IW_DMRS}. Further, we consider the sub-carrier spacing of 15kHz and FFT size of 2048 for the OFDM transmission. The gain is demonstrated under trained as well as untrained channel models and interference scenarios. The gain is quantified in terms of the SNR gap and complexity of whitening. The SNR gap is defined as 
\begin{align}
\nonumber \text{SNR gap} &= \text{SNR to achieve 10\% BLER }  -
\\ \text{SNR to} &\text{ achieve 10\% BLER for the best IW option}.
\label{eq:snr_gap}
\end{align}
Note that the low SNR gap implies low $|P_e(s;\theta) - P^{(min)}_{e}(s)|$ for $P^{(min)}_{e}(s)=0.1$.

\begin{figure}
	\vspace{-5mm}
	\centering
	\includegraphics[width=\columnwidth]{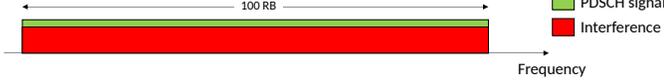}	
	\caption{{\footnotesize Untrained interference occupancy-3: full-band interference}.}
	\label{fig:int_occ3}
	\vspace{-5mm}
\end{figure}

\begin{figure}
	\centering
	\begin{subfigure}[b]{0.48\columnwidth}
		\centering
		\includegraphics[width=\columnwidth]{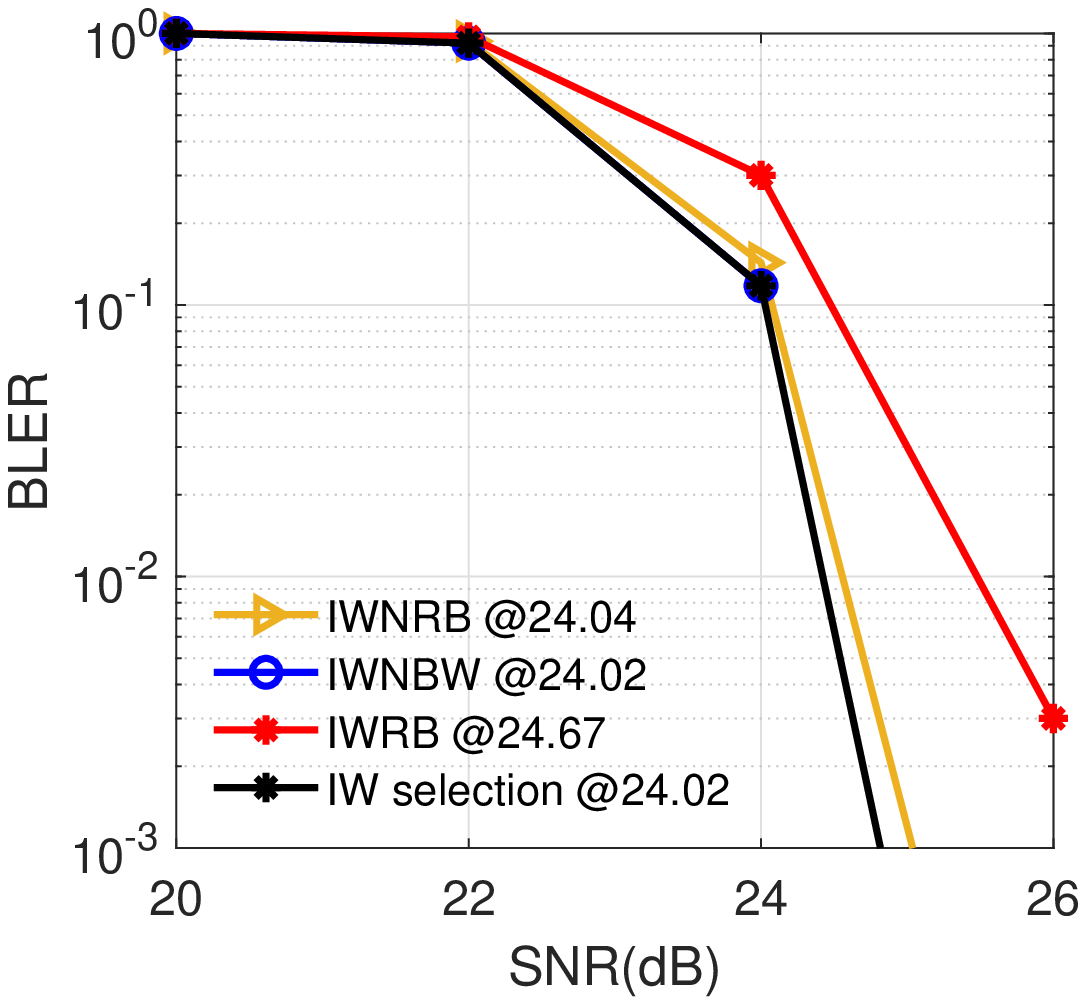}	
		\caption{{\footnotesize BLER at SIR=50dB}.}
		\label{fig:bler_mcs15_sir50_occ3}	
	\end{subfigure}
	\begin{subfigure}[b]{0.48\columnwidth}
		\centering
		\includegraphics[width=\columnwidth]{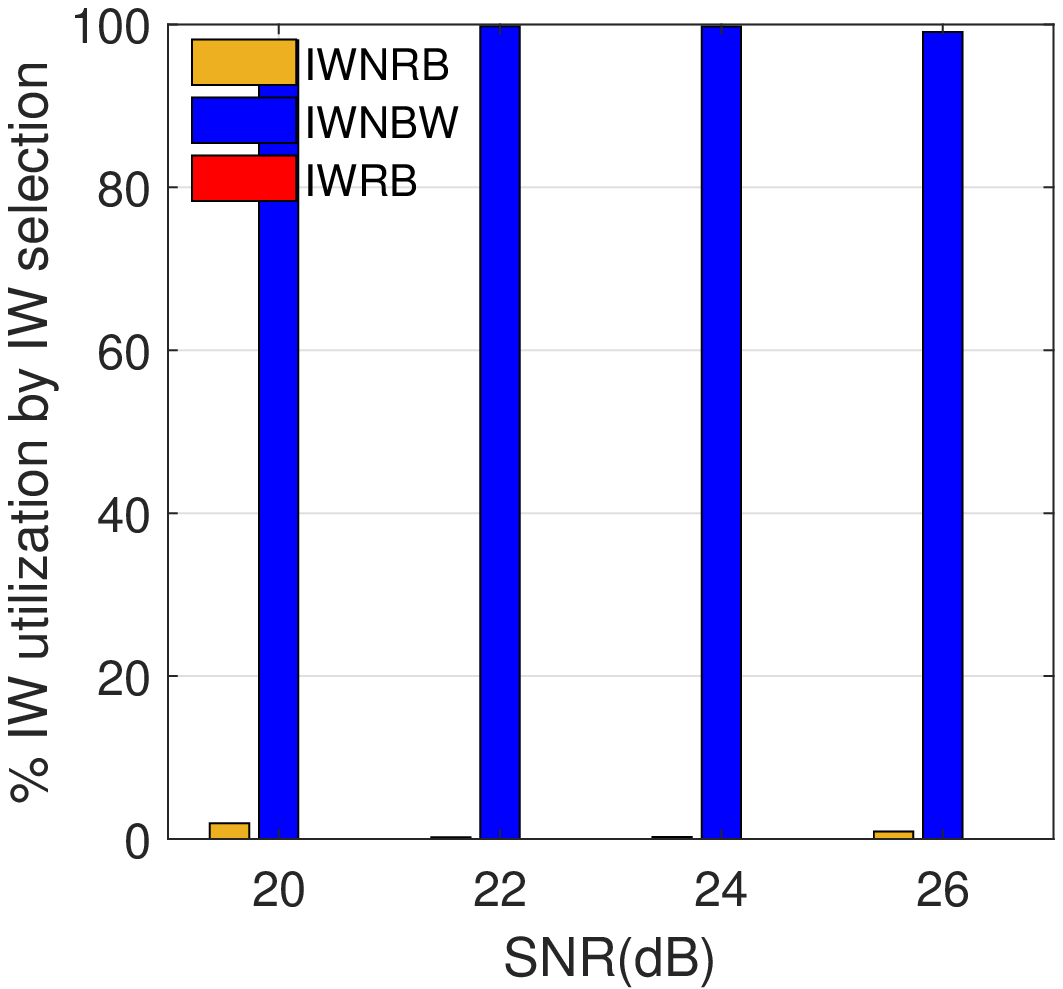}
		\caption{{\footnotesize Utilization at SIR=50dB}.}
		\label{fig:util_mcs15_sir50_occ3}
	\end{subfigure}
	\caption{\footnotesize IW selection at MCS-15 under untrained channel TDLA-30, SIR=50dB in untrained interference occupancy-3.}
	\label{fig:bler_results_int_occ3_sir50}
\end{figure}

\begin{figure*}
	\centering
	\begin{subfigure}[b]{0.24\linewidth}
		\centering
		\includegraphics[width=\columnwidth]{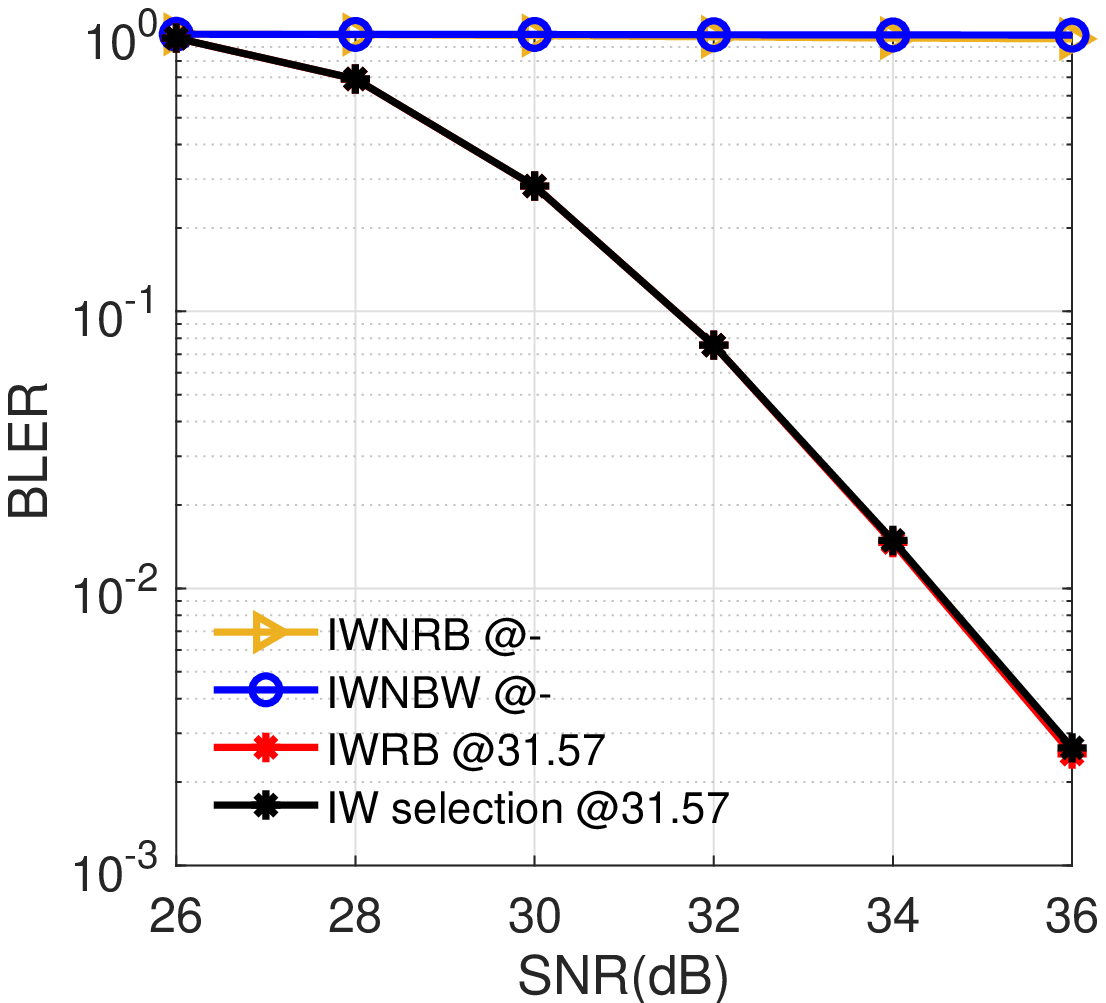}
		\caption{{\footnotesize BLER at SIR=0dB}.}
		\label{fig:bler_mcs15_sir0_occ3}
	\end{subfigure}
	\begin{subfigure}[b]{0.24\linewidth}
		\centering
		\includegraphics[width=\columnwidth]{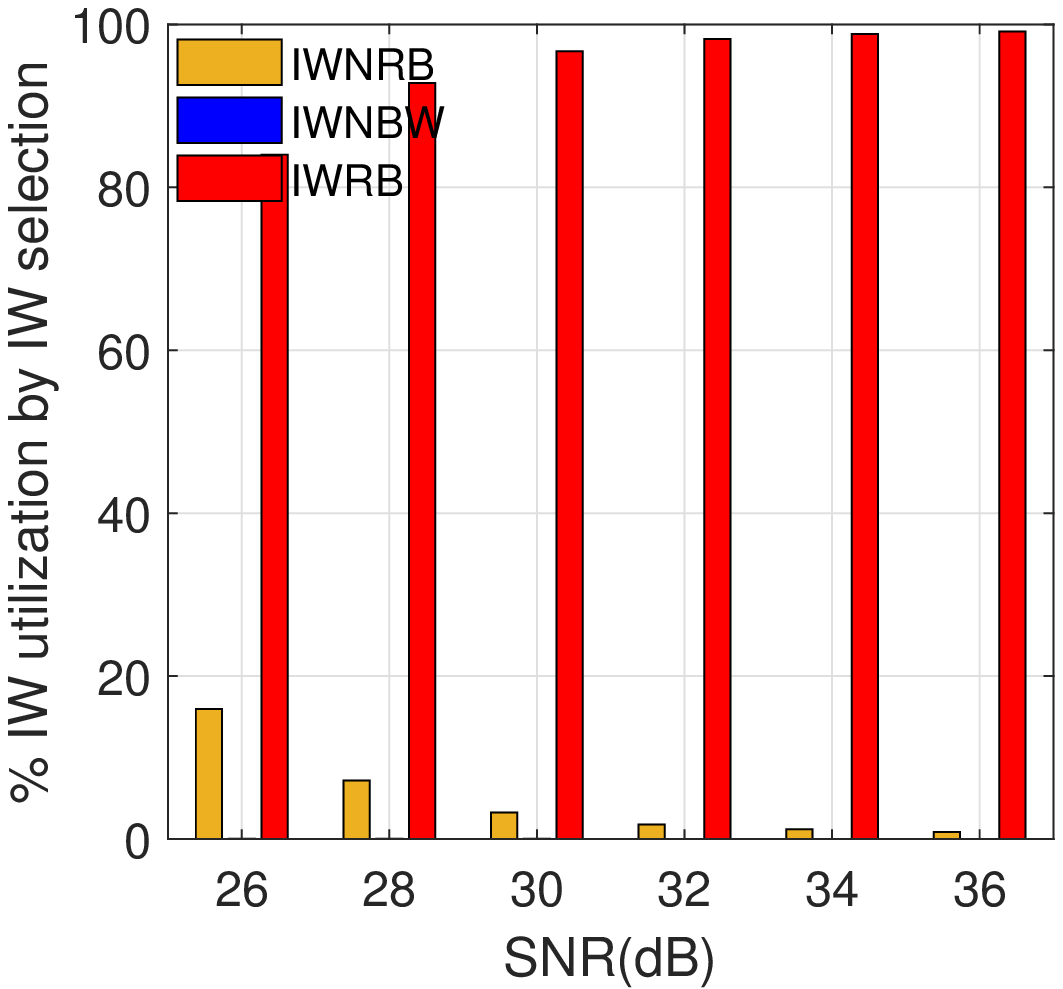}
		\caption{{\footnotesize Utilization at SIR=0dB}.}
		\label{fig:util_mcs15_sir0_occ3}
	\end{subfigure}
	\begin{subfigure}[b]{0.24\linewidth}
		\centering
		\includegraphics[width=\columnwidth]{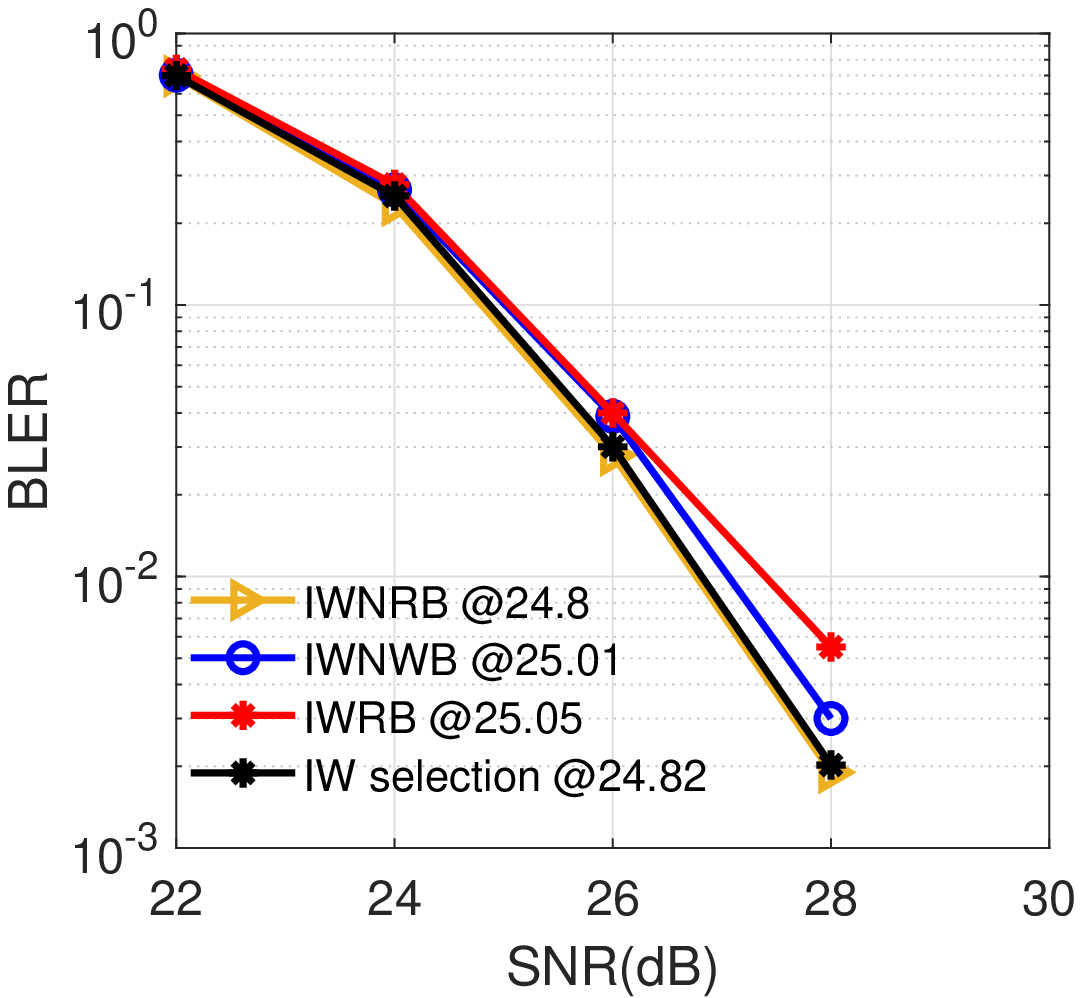}
		\caption{{\footnotesize BLER at SIR=20dB}.}
		\label{fig:bler_mcs15_sir20_occ3}
	\end{subfigure}
	\begin{subfigure}[b]{0.24\linewidth}
		\centering
		\includegraphics[width=\columnwidth]{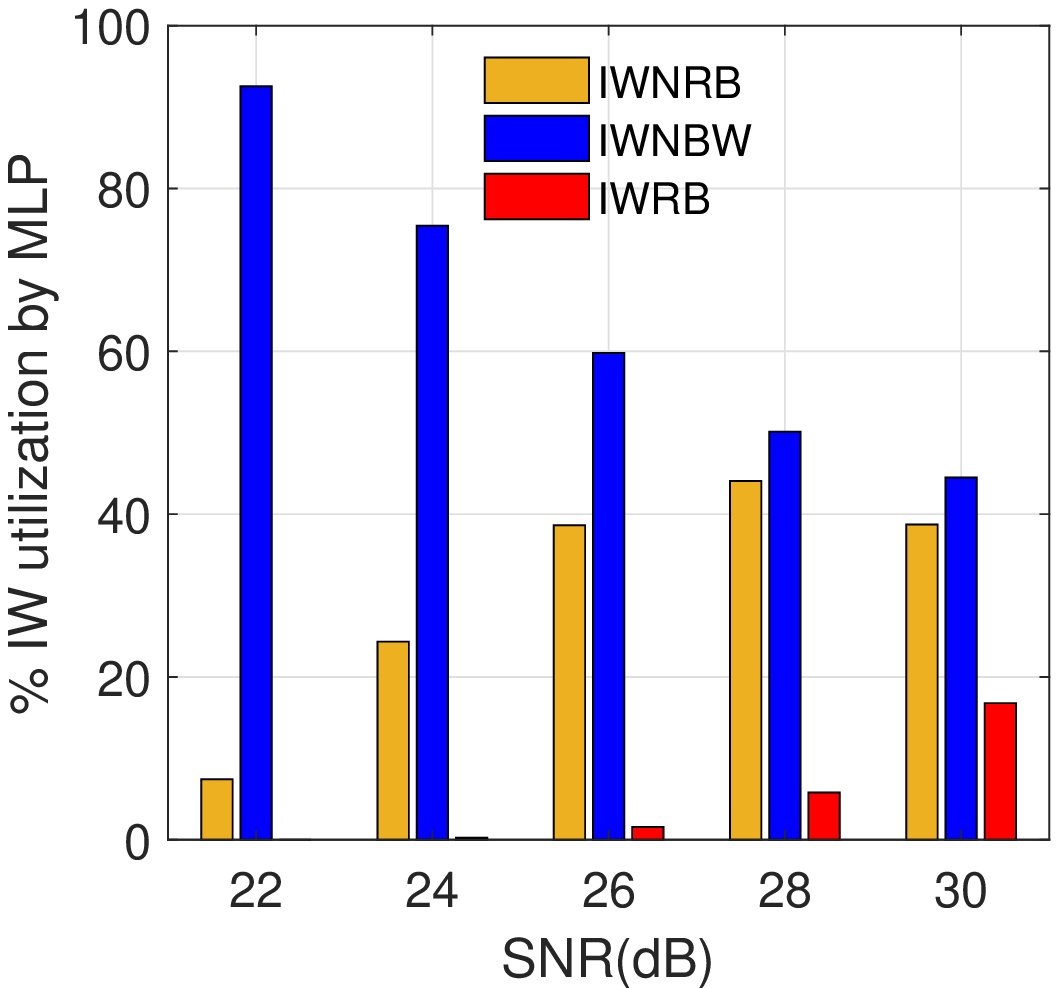}
		\caption{{\footnotesize Utilization at SIR=20dB}.}
		\label{fig:util_mcs15_sir20_occ3}
	\end{subfigure}
	\caption{\footnotesize IW selection at MCS-15 under untrained channel TDLA-30, SIR=0dB and 20dB in untrained interference occupancy-3.}
	\label{fig:bler_results_int_occ3}
\end{figure*}

In Fig. \ref{fig:bler_results_int_occ3}, we show the BLER under SIR $=0$ and $20$dB along with utilization of IW options by the proposed selection algorithm under an untrained interference occupancy-3 and untrained channel model of tapped delay line A with 30Hz Doppler frequency (TDLA-30). The interference occupancy-3 includes a full-band interference occupying entire bandwidth of the signal as  depicted in Fig. \ref{fig:int_occ3}. The BLER plots Fig. \ref{fig:bler_results_int_occ3} in also show SNR required to achieved 10\% BLER under each algorithm after the @ symbol in the legend. 

At SIR $=0$dB, the best IW option is IWRB as it achieves 10\% BLER at $31.57$dB SNR as seen in Fig \ref{fig:bler_mcs15_sir0_occ3}. The proposed IW selection algorithm also achieves 10\% BLER at 31.57dB SNR. Thus, the SNR gap for the selection algorithm 0dB. The complexity of interference whitening with the proposed method is $\mathcal{O}(N^3)$ at SIR $=0$dB since it utilizes IWRB option as shown in Fig. \ref{fig:util_mcs15_sir0_occ3}. At low SIR, the BLER with IWNBW does not drop below 10\% even at high SNR as seen in Fig. \ref{fig:bler_mcs15_sir0_occ3}. Therefore, the SNR gap for IWNBW at low SIR is $\infty$.

\begin{table*}
	\centering
	\setcellgapes{1pt}\makegapedcells
	\caption{\footnotesize{Performance of proposed IW selection under untrained interference occupancy-3}}
	\begin{tabular}{|c|c|c|c|c|c|c|c|c|c|c|}
		\hline
		& &\multicolumn{3}{c|} {Proposed IW selection} &\multicolumn{3}{c|}{IWRB}  &\multicolumn{3}{c|} {IWNBW}\\
		\hline
		& &\multicolumn{2}{c|} {SNR gap (dB)} 		& {Complexity} &\multicolumn{2}{c|} {SNR gap(dB)} 		& {Complexity}	&\multicolumn{2}{c|} {SNR gap (dB)} 		& {Complexity}	\\
		\hline		
		MCS & Channel $\rightarrow$ & EPA-5 & TDLA-30 & EPA-5 \& 	& EPA-5 & TDLA-30 	 & EPA-5 \& & EPA-5 & TDLA-30 	 & EPA-5 \&\\		
		& SIR (dB)$\downarrow$ &		&(untrained)&	 TDLA-30 	&		& (untrained)& TDLA-30 &	& (untrained)& TDLA-30 \\
		\hline
		MCS-0 			&0	 &0	 	&0 	&$\mathcal{O}(N^3)$ &0 		&0 	&  &$\infty$	&$\infty$	&\\
		\cline{2-7}
		\cline{9-10}
		&50  &0		&0	&$\mathcal{O}(N)$	&0.60	& 0.72&  &0	&0	&	\\
		\cline{1-7}
		\cline{9-10}
		MCS-7 			& 0  &0	 	&0	&$\mathcal{O}(N^3)$ &0 		&0 	&    &$\infty$	&$\infty$	& \\
		\cline{2-7}
		\cline{9-10}
		&50  &0.02	&0.01&$\mathcal{O}(N)$	&0.47	&0.62&   &0	&0	&	\\
		\cline{1-7}
		\cline{9-10}
		MCS-19 			&20  &0	 	& 0	&$\mathcal{O}(N^3)$ &0		&0 	&  $\mathcal{O}(N^3)$   &$\infty$	&$\infty$	&  $\mathcal{O}(N)$ \\
		\cline{2-7}
		\cline{9-10}
		&50  &0		&0.06	&$\mathcal{O}(N)$	&0.11&0.42&  &0	&0	&	\\
		\cline{1-7}
		\cline{9-10}
		MCS-27 			&30   &0 	&0 	&$\mathcal{O}(N^3)$ &0 		&0 	&    &$\infty$	&$\infty$	& \\
		\cline{2-7}
		\cline{9-10}
		&50  &0.01	&0.03&$\mathcal{O}(N)$&0.60		&0.63&  &0	&0	&\\
		\hline
	\end{tabular}
	\label{table:result_int_occ3}
	\vspace{-2mm}
\end{table*}

At SIR $=50$dB, the best IW option is IWNBW as it achieves 10\% BLER at 24.02dB SNR as seen in Fig \ref{fig:bler_mcs15_sir50_occ3}. The proposed IW selection algorithm also achieves 10\% BLER at $24.02$dB SNR. Thus, the SNR gap for the selection algorithm is 0dB. The complexity of the selection method is $\mathcal{O}(N)$ since it utilizes IWNBW as shown in Fig. \ref{fig:util_mcs15_sir50_occ3}. The SNR gap for IWRB is $24.67-24.02=0.65$dB. Further, the complexity of the IWRB is $\mathcal{O}(N^3)$. At high SNR, the IWNBW is the best IW option. Therefore, SNR gap for IWNBW is $0$dB. Finally, in Fig. \ref{fig:bler_mcs15_sir20_occ3} and \ref{fig:util_mcs15_sir20_occ3}, we can see that the proposed method appropriately selects an IW option to achieve the lowest BLER at 20dB SIR. 

A more extensive comparison between the IW selection algorithm, IWRB, and IWNBW is shown in Table \ref{table:result_int_occ3} for untrained interference occupancy-3. At low SIR, the best IW option is IWRB and the IW selection achieves same BLER as IWRB resulting in SNR gap of 0dB. Since the IW selection utilizes IWRB option at low SIR, the complexity of whitening is $\mathcal{O}(N^3)$. At high SIR, the SNR gap for IW selection is below $0.05$dB indicating the proposed algorithm achieves BLER very similar to the best IW option of IWNBW. Further, the complexity of whitening with the proposed algorithm is $\mathcal{O}(N)$. On the other hand, IWRB has higher SNR gap and requires higher complexity of $\mathcal{O}(N^3)$.

From the results, we can see that unlike any candidate IW option, the proposed IW selection method achieves minimum BLER in all scenarios. Further, it also has lower computational complexity than IWRB at noise-dominant scenarios (high SIR) and has significantly better BLER performance than IWNBW at interference-dominant scenarios (low SIR).

\section{Conclusion}
\label{sec:Conclusion}
In this paper, we proposed a supervised learning based algorithm to select an IW option to compute the interference-plus-noise covariance matrix from sparsely located DMRS. The proposed algorithm employs a single neural network and selects the best IW option to achieve minimum BLER under trained as well as untrained scenarios of interference occupancy, SIR, SNR, and MCS. Results also show that the proposed algorithm reduces the complexity of whitening from $\mathcal{O}(N^3)$ to $\mathcal{O}(N)$ in scenarios with spatially uncorrelated interference-plus-noise. The proposed method does not require any prior knowledge of the interference occupancy or power. Further, this method utilizes information extracted from current time slot only. Therefore, it is applicable in scenarios where CCI statistics change in each time slot.

\bibliographystyle{IEEEtran}
\bibliography{IEEEabrv,SBPreference}

\end{document}